\documentclass[revtex,amsmath,amssymb,superscriptaddress,reprint]{revtex4-2}
\usepackage[dvipsnames]{xcolor}
\usepackage[pagebackref=false]{hyperref}
\hypersetup{
    colorlinks  =   true,
    linkcolor   =   red,
    citecolor   =   blue,
    filecolor   =   magenta,
    urlcolor    =   magenta}
\usepackage{graphics,amssymb,amsmath,epsfig,color}
\usepackage{graphicx}
\usepackage{dcolumn}
\usepackage{bm}
\usepackage{color}
\usepackage{ulem}
\usepackage{siunitx}

\begin{document}

\title{Light-Induced Microwave Noise in Superconducting Microwave-Optical Transducers} 

\author{Mingrui Xu}
\thanks{These authors contributed equally to this work.}
\affiliation{Department of Electrical Engineering, Yale University, New Haven, Connecticut 06520, USA}

\author{Chunzhen Li}
\thanks{These authors contributed equally to this work.}
\affiliation{Department of Electrical Engineering, Yale University, New Haven, Connecticut 06520, USA}

\author{Yuntao Xu}
\affiliation{Department of Electrical Engineering, Yale University, New Haven, Connecticut 06520, USA}

\author{Hong X. Tang}
\email{hong.tang@yale.edu}
\affiliation{Department of Electrical Engineering, Yale University, New Haven, Connecticut 06520, USA}


\begin{abstract}
Microwave-to-optical transducers are integral to the future of superconducting quantum computing, as they would enable scaling and long-distance communication of superconducting quantum processors through optical fiber links. However, optically-induced microwave noise poses a significant challenge in achieving quantum transduction between microwave and optical frequencies. In this work, we study light-induced microwave noise in an integrated electro-optical transducer harnessing Pockels effect of thin film lithium niobate. We reveal three sources of added noise with distinctive time constants ranging from sub-100~nanoseconds to milliseconds. Our results gain insights into the mechanisms and corresponding mitigation strategies for light-induced microwave noise in superconducting microwave-optical transducers, and pave the way towards realizing the ultimate goal of quantum transduction.
\end{abstract}

\maketitle
\section{Introduction}
The microwave-to-optical (MO) transducer plays an essential role in future quantum networks \cite{schoelkopf2008wiring,zhong2020proposal,jiang2007distributed,cirac1997quantum}, where the quantum information is processed by superconducting qubits \cite{clarke2008superconducting,devoret2013superconducting} at microwave frequencies and transmitted using optical photons \cite{o2009photonic}. The added noise is a critical performance parameter for MO transducers, as any excessive noise would deteriorate the fidelity of the signal transduction. Typically, for a GHz electro-optical (EO) transducer \cite{tsang2010cavity, tsang2011cavity} operating at low input powers, simple refrigeration of the device to tens of milli-kelvin temperature is sufficient to suppress thermal excitations in the microwave resonators. However, maintaining the microwave resonator at the ground state while retaining high transduction efficiency in the presence of a high-intensity optical drive is particularly challenging. Various schemes have been implemented to realize transduction at microwave ground state, including cavity EO \cite{fu2021cavity,hease2020bidirectional}, rare-earth ion \cite{rochman2023microwave} and piezo-optomechanics \cite{mirhosseini2020superconducting,forsch2020microwave} systems. However, the conversion efficiencies demonstrated thus far still fall significantly below unity. This fact hinders the realization of quantum state transduction, where an intense optical drive is critically required to achieve high conversion efficiency \cite{zeuthen2020figures}.

Recently, pulsed optical drive has emerged as a promising approach to achieve higher conversion efficiency in various MO transduction schemes \cite{sahu2022quantum, fu2021cavity, han2020cavity, forsch2020microwave, mirhosseini2020superconducting}. This technique allows for high-power optical drive and reduces overall heat dissipation at the same time. Near-unity cooperativity has been achieved in a bulk lithium niobate EO platform, utilizing Watt-scale pulses \cite{sahu2022quantum}. However, at milli-Kelvin temperature where the cooling power is relatively weak, a strong optical drive applied to integrated transducers results in significant heating up of the device. This heat introduces additional microwave noise through photon absorption by dielectrics \cite{mobassem2021thermal} and superconductor \cite{il2000picosecond,day2003broadband,beck2011energy,kardakova2013electron}. Therefore, there is a pressing need to gain a fundamental understanding of the mechanism underlying light-induced microwave noise in superconducting MO transducers. Such understanding is crucial to explore potential strategies for noise suppression.

In this work, we study the microwave noise induced by pulsed optical drive in an integrated superconducting-EO transducer. The transducer is implemented on a thin-film lithium niobate (TFLN)-niobium nitride (NbN) hybrid material platform as introduced in reference \cite{xu2022light} and displayed in Fig.~\ref{method}(a). Instead of investigating the microwave noise as a whole, which has been done in our previous work \cite{fu2021cavity}, three sources of noise have been identified here, as illustrated in Fig.~\ref{method}(b) and (d), including heating of the external thermal bath $n_\mathrm{bg}$, fast time-scale heating and cooling of an intrinsic thermal bath during pulse rise and fall time $n_\mathrm{i,fast}$, and slow time-scale heating and cooling of the intrinsic thermal bath during pulse-on and -off time $n_\mathrm{i,slow}$. Moreover, the corresponding time constants for each noise source are measured and analyzed. $n_\mathrm{bg}$ has a relaxation time much longer than 5~ms, while $n_\mathrm{i,fast}$'s is much shorter than 100~ns. The time constant of $n_\mathrm{i,slow}$ is more complicated, containing components ranging from 33\,{\textmu}s to 6~ms.
We also explore the dependence of light-induced microwave noise on pulse peak power, pulse width and pulse period. Both $n_\mathrm{bg}$ and $n_\mathrm{i,slow}$ are related to average power, while $n_\mathrm{i,fast}$ can only be affected by pulse peak power. These results suggest different noise generation mechanisms in the system. Lastly, we discuss potential strategies to mitigate light-induced microwave noise and provide insights for the design of future integrated EO transducers, with the aim of further suppressing these detrimental noise effects.

\section{Method}

\begin{figure*}[ht]
\centering
\includegraphics[width=0.8\textwidth]{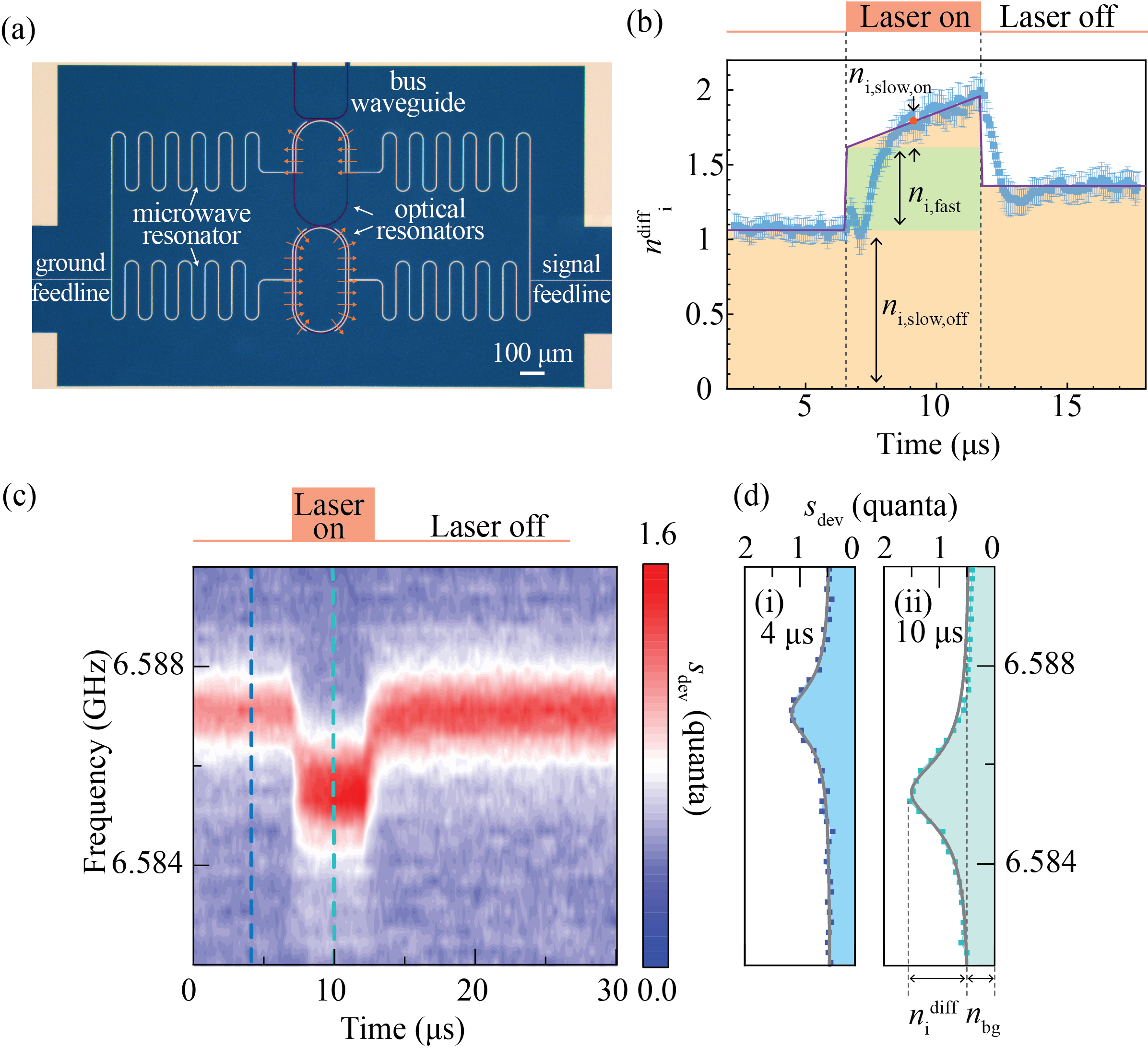}
\caption{\label{method} 
(a) False-colored microscopic image of the EO transducer. Two strongly coupled optical racetrack resonators are integrated with a superconducting microwave resonator. The microwave
resonator is linked to a coplanar waveguide (CPW, not shown) for both RF coupling and DC biasing. The electrical field direction of the designed microwave mode is marked in the coupling regime. The figure is taken from reference\cite{xu2022light}. (b) is the temporal evolution of $n_\mathrm{i}^\mathrm{diff}$ obtained at a pulse peak power of 6~dBm, the pulse width of 5\,{\textmu}s and the pulse period of 1~ms. The optical drive is turned on and off at approximately 6.6\,{\textmu}s and 11.8\,{\textmu}s. Solid purple line is the extrapolation of  $n_\mathrm{i}^\mathrm{diff}$. The heat map (c) shows the time evolution of the noise spectrum with a pulse peak power of 5~dBm, the pulse width of 5\,{\textmu}s and the pulse period of 1~ms. (d)(i) and (ii) show the snapshots of power spectral density when the pulsed optical drive is on and off. Data in (d)(i) and (ii) are cross-sections of (c) marked in correspondingly-colored dashed lines. }
\end{figure*}

To investigate the dynamics of thermal baths temperatures during a pulsed optical drive, we employ a calibration schedule to determine the input line attenuation by using a variable temperature stage (VTS). We subsequently record the time evolution of the noise power spectrum using an ultra high frequency lock-in amplifier. This process involves two steps. Firstly, we select a specific frequency and measure the noise power within a time window relative to the optical drive. We repeat this process numerous times and average the results to obtain a time trace. Secondly, we sweep the frequency and repeat the previous step for each frequency of interest. We alternate the measurement with the optical drive turning on and off. The difference between the data of on-off experiments yields the excessive noise spectrum $s_\mathrm{dev}$ heat map (Fig.~\ref{method}(c)) from the EO transducer.

The dynamic noise measurement protocol outlined above requires precise knowledge of the state of the resonance, including the coupling rates and resonant frequency, which can vary with time. We achieve this by probing the resonator with a weak coherent tone to monitor the time evolution of the resonance. This measurement is similar to acquiring a heat map of noise power, but instead of recording noise power, we record the complex reflection of the resonator. By analyzing the reflection spectrum obtained at different time points, we can fit it to a Lorentzian function to extract the coupling rates and resonant frequency accurately. \par

In measuring the time evolution of noise spectrum, there is an important trade-off between the time resolution and frequency resolution. On one hand, studying the transition of optically induced microwave noise on the scale that is sub-$100\,\mathrm{ns}$ is a challenging task. It is mainly due to two limiting factors: (a) low-pass filter settling time of the lock-in amplifier. (b) the rise and fall time of the pulsed drive (the acousto-optic modulator), which is on the scale of $35\,\mathrm{ns}$. While the factor (b) is fixed by our physical setup, we maximize the bandwidth for the low-pass filter in the lock-in amplifier, with a time constant of $t_\mathrm{c} = 30~\mathrm{ns}$, to record the noise spectrum with the best resolution limited by our electronics as shown in Fig.~\ref{n_dynamics}(c).

On the other hand, while resolving the frequency envelope of the noise spectrum, we use $t_c=1$\,{\textmu}s, which limits the temporal resolution to microsecond scale. Fig.~\ref{method}(c) and (d) is an example. As shown in Fig.~\ref{method}(c) and (d), a consistent background noise of 0.55 quanta is observed over the frequency range of interest. In the pulse-off period (Fig.~\ref{method}(d)(i)), there is a resonant peak at 6.587~GHz associated with finite thermal excitation. When the pulse is on (Fig.~\ref{method}(d)(ii)), this resonance down-shifts by 1.5~MHz, and the resonant linewidth broadens. At the new resonant frequency, there is stronger thermal excitation. The thermal bath occupancy can be extracted by applying the model described in the Appendix, where $n_\mathrm{bg}$ and $n_\mathrm{i}^\mathrm{diff}$ are the only free parameters as shown in Fig.~\ref{method}(d)(ii). In the next section, the dynamics of both will be investigated. 

\section{Results}

\subsection{Dynamics of $n_\mathrm{bg}$}

\begin{figure*}[ht]
\centering
\includegraphics[width=0.8\textwidth]{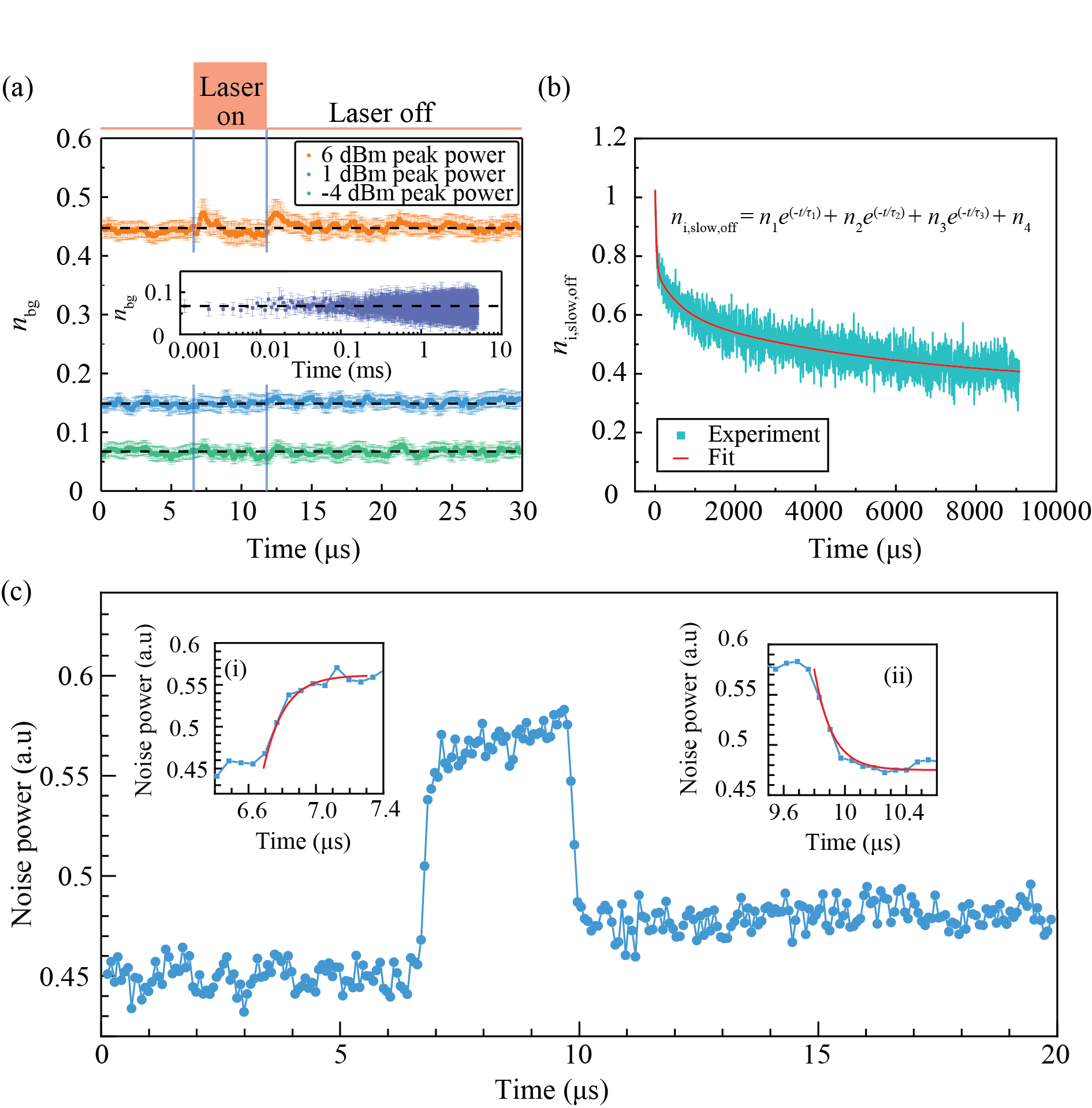}
\caption{
\label{n_dynamics}
(a) Dynamics of $n_\mathrm{bg}$ over 30\,{\textmu}s, the optical drive is turned on and off at approximately 6.6\,{\textmu}s and 11.8\,{\textmu}s as marked. Pulsed drive is configured with different peak powers and the same duty cycle of 0.5\%. The inset shows the dynamics of $n_\mathrm{bg}$ over a time span of 5~ms after the optical drive is turned off. The pulsed drive is configured with peak power of 0~dBm and pulse width of 10\,{\textmu}s. Black dashed lines are all horizontal lines as guide to the eye. (b) Time traces of $n_\mathrm{i,slow,off}$ while peak power is 0~dBm, pulse width is 10\,{\textmu}s, and pulse period is 10~ms. The cyan trace is the experimental data, while the red trace is the fitting. The decay curve can be fitted to a triple exponential decay function showing three time constants, where $\tau_1 = 33\pm6$\,{\textmu}s, $\tau_2 = 0.6\pm 0.1~\mathrm{ms}$ and $\tau_3 = 6\pm1~\mathrm{ms}$. (c) Output noise power time dynamics with $t_c = 30~\mathrm{ns}$. (i) and (ii) are the zoom-in views of the rise and fall edges of (c), where fits to exponential functions are shown in red curves.
}
\end{figure*}

We first examine $n_\mathrm{bg}$, which combines the thermal excitation due to the external thermal bath of the superconducting resonator and the added noise from the output readout line. By analyzing the heat map shown in Fig.~\ref{method}(c) that captures the evolution of noise power spectrum over time, it becomes evident that $n_\mathrm{bg}$ remains constant over time on the scale of microseconds and exhibits no discernible frequency dependence within the 10~MHz span.

Fig.~\ref{n_dynamics}(a) illustrates the dynamics of $n_\mathrm{bg}$. The data points are collected while the optical drive is on and off, and different traces correspond to different optical drive peak powers. It is evident that in all these traces, $n_\mathrm{bg}$ does not change with regard to the status of the optical drive being on or off. Due to the finite time constant (1\,{\textmu}s) of the filter in our measurement electronics, the noise spectrum is artificially distorted right after the rise and fall edges of the optical drive. These small artificial bumps can be observed in the orange trace 

To investigate $n_\mathrm{bg}$ on a longer time scale, we characterized its relaxation between optical drive pulses with a period of 5~ms, as shown in Fig.~\ref{n_dynamics}(a) inset. In comparison to the dashed horizontal lines provided as visual guides, we find that $n_\mathrm{bg}$ does not change significantly within the millisecond time scale, indicating that $n_\mathrm{bg}$ likely has a relaxation time much longer than 5~ms. These observations rule out the possibility of $n_\mathrm{bg}$ being caused by leaked infrared light impacting superconducting components, such as JTWPA, in the cryogenic microwave circuits. Otherwise, we would observe more changes in $n_\mathrm{bg}$ while the optical drive is turned on and off. Instead, the results suggest that the heating of the dilution refrigerator might be the cause of the excessive $n_\mathrm{bg}$.

To gain further understanding of the source of $n_\mathrm{bg}$, we characterize the dependence of $n_\mathrm{bg}$ on the average drive power ($P_\mathrm{avg} = P_\mathrm{peak} \times \mathrm{duty cycle}$) in Fig.~\ref{n_Pdependence}(a). This plot aggregates three sets of measurement results where we sweep the pulsed drive peak power, pulse width or pulse period. By fitting all data to an exponential growth function (dashed red line), we found that $n_\mathrm{bg} \approx P_\mathrm{avg}^{0.82}$. The fact that all data points from three different measurement configurations coalesce very well also indicates that $n_\mathrm{bg}$ is only affected by average optical drive power.
These results confirm that $n_\mathrm{bg}$ is a result of the heating of the dilution refrigerator.

\subsection{Dynamics of $n_\mathrm{i}^\mathrm{diff}$}

\subsubsection{``Fast" and ``Slow" responding noise components}

In great contrast to $n_\mathrm{bg}$, which seems to have a very long relaxation time and does not respond to different illumination conditions, the dynamics of $n_\mathrm{i}^\mathrm{diff}$ exhibits a significantly richer behavior across various time scales. In order to further understand the mechanisms of these noise components, we separate $n_\mathrm{i}^\mathrm{diff}$ into three components:
\begin{equation}
    n_\mathrm{i}^\mathrm{diff} = n_\mathrm{i,fast} + n_\mathrm{i,slow,off} + n_\mathrm{i,slow,on}.
\end{equation}
This division is also illustrated in Fig.~\ref{method}(b), where an example time trace of $n_\mathrm{i}^\mathrm{diff}$ is presented. To account for the slow transition slope masked by the low-pass filter, we extrapolate the data (solid purple line) based on the linear fit of steady measurement results.

As shown in Fig.~\ref{method}(b), we define $n_\mathrm{i,fast}$ as the ``fast-responding'' part of the $n_\mathrm{i}^\mathrm{diff}$. It is obtained from the time evolution traces of $n_\mathrm{i}^\mathrm{diff}$ as $n_\mathrm{i,fast} = n_\mathrm{i}^\mathrm{diff}|_\mathrm{after laser}-n_\mathrm{i,slow,off}$. Here  $n_\mathrm{i}^\mathrm{diff}|_\mathrm{after laser}$ is defined as $n_\mathrm{i}^\mathrm{diff}$ at the moment immediately after the optical drive is turned on. By assuming a linear change in $n_\mathrm{i}^\mathrm{diff}$ when the optical drive is on, we extrapolate $n_\mathrm{i}^\mathrm{diff}|_\mathrm{after laser}$ from the steady state measurement results. The time when the optical drive is turned on and off are determined by the moment when the resonant frequency shifts in the coherent measurement with $\tau_c = 30~\mathrm{ns}$. 

The other two noise components are the ``slow-responding'' part of $n_\mathrm{i}^\mathrm{diff}$: $n_\mathrm{i,slow,off}$ is defined as the $n_\mathrm{i}^\mathrm{diff}$ right before the optical drive is on; $n_\mathrm{i,slow,on}$ is the buildup of  $n_\mathrm{i}^\mathrm{diff}$ during the pulse-on period. Given the narrow optical drive pulse width, it is reasonable to assume the increase of $n_\mathrm{i}^\mathrm{diff}$ is linear within the first 5\,{\textmu}s after the optical drive is turned on, and this is the basis for the linear extrapolation to obtain $n_\mathrm{i}^\mathrm{diff}$ during the pulse-on period. We use $n_\mathrm{i}^\mathrm{diff}$ at the middle point of the optical drive pulse to indicate the average intrinsic thermal bath temperature. Therefore, we define $n_\mathrm{i,slow,on} = n_\mathrm{i}^\mathrm{diff}|_\mathrm{midpulse} - n_\mathrm{i}^\mathrm{diff}|_\mathrm{after laser}$.

\subsubsection{``Fast" light induced microwave noise generation}

\begin{figure*}[ht]
\centering
\includegraphics[width=0.8\textwidth]{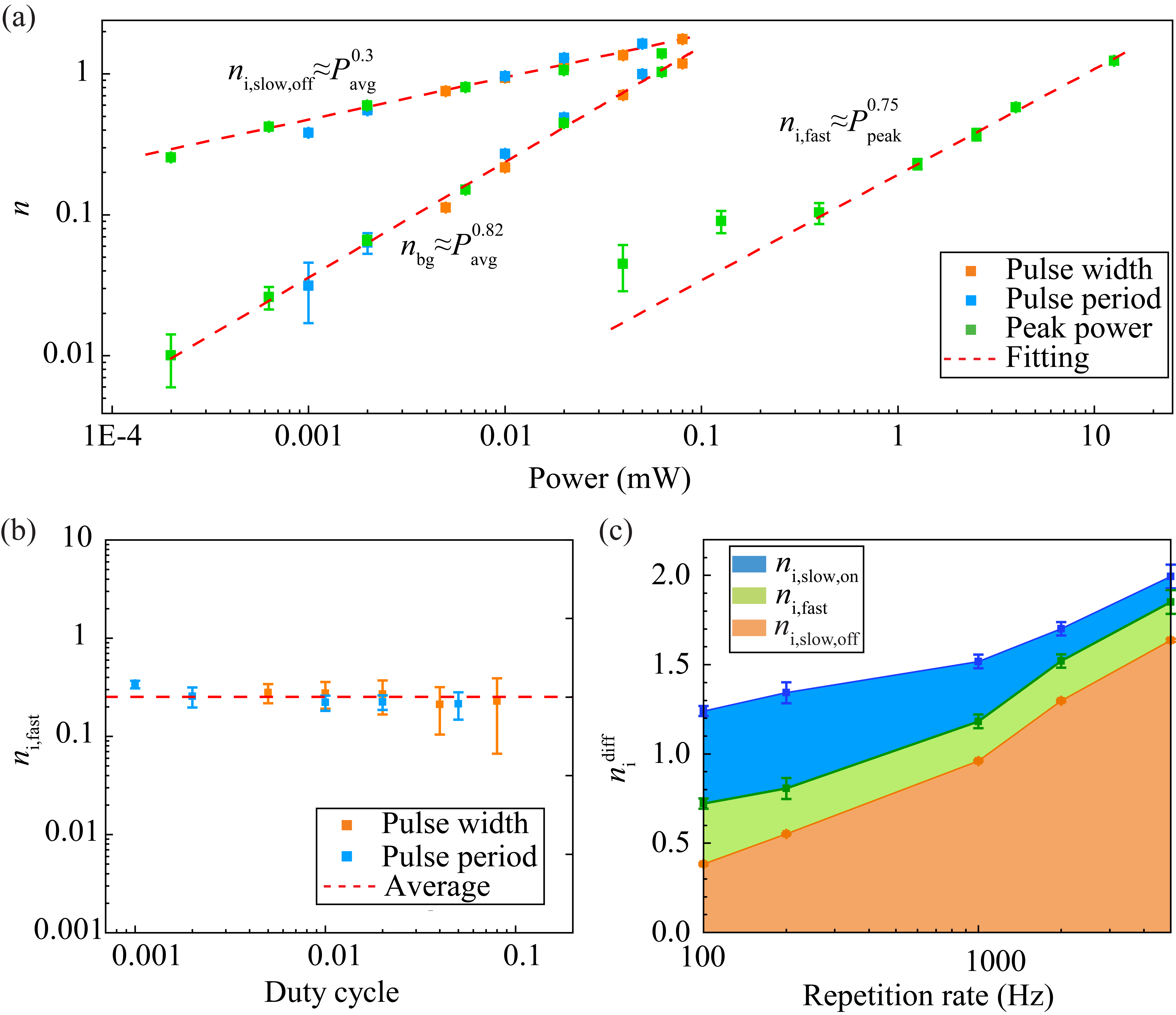} 
\caption{
\label{n_Pdependence}
(a)$n_\mathrm{bg}$ and $n_\mathrm{i,slow,off}$ as a function of average power of the optical drive and $n_\mathrm{i,fast}$ as a function of peak power of the optical drive. Orange data points corresponds to optical drives with 0~dBm peak power, 1~ms period, while pulse width is swept from 5\,{\textmu}s to 80\,{\textmu}s. Blue data points corresponds optical drives with 0~dBm peak power, 10\,{\textmu}s pulse width, while the period is swept from 0.2~ms to 10~ms. Green data points corresponds to optical drives with $5$\,{\textmu}s pulse width and 1~ms pulse period, while peak power is swept from -9~dBm to 11~dBm. Dashed red line is an exponential fit to all the data. The fittings show $n_\mathrm{bg} \approx P_\mathrm{avg}^{0.82}$, $n_\mathrm{i,fast} \approx P_\mathrm{peak}^{0.75}$ and $n_\mathrm{bg} \approx P_\mathrm{avg}^{0.3}$. 
(b) shows dependence of $n_\mathrm{i,fast}$ on the duty cycle of optical drives. Orange and Blue data points correspond to the same pulse configuration as those in (a). Red dashed line is the average value of the data.
(c) Dependence of $n_\mathrm{i}^\mathrm{diff}$ and its different component on the repetition rate of optical drives. Pulse peak power is fixed at 0~dBm, pulse width is 10\,{\textmu}s, and repetition rate is swept from 100~Hz to 5000~Hz.
}
\end{figure*}

In Fig.~\ref{n_dynamics}(c), we show the dynamics of $n_\mathrm{i}^\mathrm{diff}$ from a high-time-resolution measurement. Because of the small time constant, the noise power measurement bandwidth is much larger than the resonance linewidth, making it difficult to accurately extract occupancy of different thermal bath. Therefore, we present the output noise power as an indication of the changes of $n_\mathrm{i}^\mathrm{diff}$.
Fits to the rise and fall edges as shown in Fig.~\ref{n_dynamics}(i) and (ii) show a relaxation time of 120~ns $\pm$ 30~ns consistently, which is mainly limited by the pulse rise and fall time. The actual transition should be much faster than 120~ns.\par

To find out the dependence of $n_\mathrm{i,fast}$ on the optical drive power, we first sweep the pulse peak power while fixing the pulse width and pulse period at 5\,{\textmu}s and 1~ms. The results are shown in Fig.\,\ref{n_Pdependence}(a), where we find $n_\mathrm{i,fast}$ increase as $P_\mathrm{peak}^{0.75}$. In order to maintain $n_\mathrm{i,fast}$ below 0.1, $P_\mathrm{peak}$ needs to be set below $0.3~\mathrm{mW}$, while the threshold for $n_\mathrm{i,fast} = 1$ is $P_\mathrm{peak} = $10~mW. 
Then, we investigate the dependence of $n_\mathrm{i,fast}$ on the pulse width and pulse period, respectively. Results in Fig.~\ref{n_Pdependence}(b) inset show $n_\mathrm{i,fast}$ does not change significantly to different pulse configurations as long as the peak power is constant. Deviations from the average value as shown in the dashed red line could be a result of different device temperatures, as a higher duty cycle leads to more average power dissipation and higher local temperature of the device, and vice versa.

This fast light-induced microwave noise could be attributed to light absorption by the superconductor \cite{il2000picosecond, day2003broadband,beck2011energy,kardakova2013electron}. It can be understood as the same mechanism as superconducting single photon detectors. When optical photons are directly absorbed by the superconductor, quasiparticles are generated in an energy down-conversion process. In materials such as NbN, this process could be much faster than 1~ns. The other possible mechanism is the local heating due to optical photons being absorbed by the optical waveguide and the surrounding substrate materials next to the superconducting resonator \cite{mobassem2021thermal}. These two mechanisms cannot be distinguished in our experiment.\par

\subsubsection{``Slow" light induced microwave noise generation}

In Fig.~\ref{n_dynamics}(b), we show the time evolution of $n_\mathrm{i,slow,off}$ up to 9~ms. We fit the decay curve to a triple-exponential decay model: $n_\mathrm{i,slow,off} = n_\mathrm{1} e^{(-t/\tau_1)}+n_\mathrm{2} e^{(-t/\tau_2)}+n_\mathrm{3} e^{(-t/\tau_3)}+n_\mathrm{4}$. Data and fitting are shown in cyan and red traces, respectively. Three different time constants, $\tau_1 = 33\pm6$\,{\textmu}s, $\tau_2 = 0.6\pm 0.1~\mathrm{ms}$ and $\tau_3 = 6\pm1~\mathrm{ms}$, are extracted. Additionally, the fitting shows a constant background of $n_\mathrm{4} = 0.36$, indicating that there is still a significant portion of $n_\mathrm{i,slow,off}$ that decays at a time scale that is much longer than $9~\mathrm{ms}$. For a typical pulsed measurement with repetition on the order of kHz, this results in the buildup of thermal fluctuation in EO transducer that accumulates through pulses sequences. 

If we take a look at the optical power dependence of $n_\mathrm{i,slow,off}$ as shown in Fig.~\ref{n_Pdependence}(a), all data shows a similar trend, which is represented by the dashed line as $n_\mathrm{i,slow,off}\propto P_\mathrm{avg}^{0.3}$. This trend could be explained by the increased material heat conductivity with higher temperature at cryogenic temperatures. Specifically, we know that at the low-temperature limit, LN and sapphire's thermal conductivity increases as the cubic temperature, and copper's thermal conductivity increases proportionally to the temperature. 
We consider a simple model comprised of a point heater (optical waveguide) thermalized to a large thermal reservoir (dilution refrigerator) through a solid media. Assuming the thermal conductivity of the solid media is dominated by phonons, which follows cubic law with regard to temperature, the equilibrium temperature of the heater would be proportional to the temperature to the power of 1/4.
In our experiment, the power-law relation between the temperature and average optical power dissipation has a factor of 0.3, suggesting that the source of $n_\mathrm{i,slow,off}$ could be attributed to the heat-up of the device and the cooper package while LN and sapphire might play a more dominant role in thermalizing the intrinsic thermal bath associated to  $n_\mathrm{i,slow,off}$.

How effective is decreasing repetition rate in suppressing $n_\mathrm{i}^\mathrm{diff}$? The answer can be found in Fig.~\ref{n_Pdependence}(c). Here, we fix the pulsed peak power at 0~dBm and pulse width at 10\,{\textmu}s, while sweeping the pulse repetition rate from 5000~Hz to 100~Hz. Data shows as the repetition rate decreases, $n_\mathrm{i,slow,off}$ also decreases with the average heat dissipation. However, $n_\mathrm{i,slow,on}$ increases with lower repetition, despite the same power and duration of each optical drive pulse. This could be explained by the decreased thermal conductivity in materials with lower temperature at cryogenic temperatures, leading to a faster increase in temperature. On the other hand, $n_\mathrm{i,fast}$ does not change significantly with the repetition, which is consistent with our observation in the previous section. Therefore, although reducing the repetition rate of optical drive pulses could suppress light-induced microwave noise, its effectiveness is limited by the fast-responding noise $n_\mathrm{i,fast}$ and the build-up of thermal noise while the optical drive is on $n_\mathrm{i,slow,on}$. As a result, the power handling capability of on-chip EO transducers should be boosted by improving the design of the chip and the packaging.

\section{Discussion and Conclusions}

This study reveals that the light-induced microwave noise is dominated by the dissipation of optical power in the EO transducer system and scattered photons shining on the superconductor. We argue that efficient fiber-chip interfaces, efficient light shielding,  and superfluid helium immersion could be the key to mitigating these sources of microwave noise. \par

First, an efficient fiber-chip interface could reduce the amount of optical power dissipating in the device package. Considering the efficiency of photons passing from the input fiber to the on-chip waveguide as $\eta_\mathrm{in}$ and that of photons passing from the waveguide to the output fiber as $\eta_\mathrm{out}$.  The the optical cavity's transmission amplitude is $T$. Assuming the optical power in the waveguide before the optical cavity is $P_\mathrm{t}$, then the power sent to the input fiber is $P_\mathrm{in} = P_\mathrm{t}/\eta_\mathrm{in}$.
For simplicity, we do not consider the optical power dissipating in the straight waveguide as the loss is usually much smaller (0.03~dB/cm for TFLN waveguides) than the extinction of optical resonances.
Then we can derive the total amount of power dissipated in the system as
\begin{equation}
P_\mathrm{heat} = P_\mathrm{t}\cdot \big[ \frac{1}{\eta_\mathrm{in}} -1 + |T|^2 (1 - \eta_\mathrm{out}) \big].
\end{equation}
This equation highlights the important role that $\eta_\mathrm{in}$ and $\eta_\mathrm{out}$ play in heat dissipation. Improving the fiber-to-chip efficiency from 50\% to 80\% can reduce 75\% of heat dissipation from the optical drive at most. 
Additionally, designing the optical resonances in the over-coupled or under-coupled regimes can lead to a larger $T$, resulting in more optical photons dissipating in the system through the external decay of the optical cavity.

Second, to address scattered light shining on the superconducting resonator, several design strategies can be employed. One approach is to use metal shields to block the direct propagation of photons in the substrate and the vacuum toward the superconducting resonator. For grating couplers, coating the backside of the grating coupler with metal can redirect light away from the superconducting resonator. In flip-chip bonded EO devices, coating the photonic chip with metal can shield stray light in the photonic chip substrate from shining on the inductor of the superconducting resonator. Additionally, using metal such as gold as light shields can also enhance the thermal conductivity between the device chip and the copper device packaging, helping dissipate heat.

Finally, immersing the device in superfluid helium can greatly increase the heat dissipation rate for the device~\cite{fan2018superconducting,xu_cavity-enhanced_2022}, due to the exceptional heat transport properties of superfluid helium. This approach can effectively suppress microwave noise induced by local heating~\cite{lane2020integrating,koolstra2019coupling,samkharadze2011integrated}. However, it does have two limitations. Firstly, the total amount of heat dissipation is still limited by the cooling power of the dilution refrigerator. For example, an average optical drive power of -6 dBm (250 \,{\textmu}W) will inevitably heat up the mixing chamber to 100 mK, resulting in additional thermal noise. Secondly, for any microwave noise generated through superconductor absorption of photons, the superfluid helium is unlikely to be effective. Nonetheless, superfluid helium immersion would provide valuable high-capacity cooling to allow the device to operate at ground state with higher drive powers and provide additional insights into the sources of light-induced microwave noises.

In conclusion, we present a study of added noise in an on-chip integrated EO transducer. We examine different sources of microwave noise, including both external and intrinsic bath. We discover that the external bath temperature increase is largely determined by the average optical drive power dissipation, while intrinsic bath is composed of various components with distinct relaxation times, ranging from sub-microseconds to beyond milli-seconds.  
We analyze the dependencies of each noise source and provide an understanding of potential noise generation mechanisms. 
Finally, we discuss various noise mitigation strategies for different noise sources. These insights can guide the design of future EO transducers for increasing power handling capability, making quantum transduction feasible.

\section*{Acknowledgement}
We acknowledge funding support from DOE Office of Science, National Quantum Information Science Research Centers, Co-design Center for Quantum Advantage (C2QA), Contract No. DE-SC0012704. HXT acknowledges partial funding support from National Science Foundation (NSF) through ERC Center for Quantum Networks (CQN) grant EEC-1941583 and early support from Army Research Office on the quantum transducer development (through grant number No. W911NF-18-1-0020). The JTWPAs used in this experiment are provided by IARPA and MIT Lincoln Laboratory. The authors would like to thank Dr. Yong Sun, Dr. Lauren McCabe, Mr. Kelly Woods, and Dr. Michael Rooks for assistance in device fabrication. 

\section*{Appendix A: Output line calibration protocol} 
\setcounter{figure}{0} 
\setcounter{equation}{0}
\renewcommand\thefigure{\Alph{section}A\arabic{figure}} 
\renewcommand\theequation{A\arabic{equation}}
Calibrating the gain and added noise of the output line is crucial for accurate determination of the absolute value of the light-induced noise in the device under test (DUT). The output line calibration is performed by sourcing the DUT with a variable temperature stage (VTS) in the mixing chamber of the dilution refrigerator, whose thermal occupation follows a Bose-Einstein distribution: $n_\mathrm{VTS}=1/\big(\mathrm{exp}(hf/k_\mathrm{B}T_\mathrm{VTS})-1\big)$. The noise power spectral density is then measured at the output of the dilution refrigerator using a spectrum analyzer or lock-in amplifier, after the amplification chain featuring a JTWPA. \par

The noise power spectrum $P$ is thus linked to the VTS output through:

\begin{equation}
P = BW\cdot G_\mathrm{sys}(n_\mathrm{VTS,s}\hbar \omega_\mathrm{s} +n_\mathrm{VTS,i}\hbar \omega_\mathrm{i} +n_\mathrm{q}\hbar \omega_\mathrm{s}+n_\mathrm{sys}\hbar \omega_\mathrm{s}).
\label{TWPA_calibration}
\end{equation}
In this equation, $BW$ represents the measurement resolution bandwidth, while $n_\mathrm{VTS,s}$ and $n_\mathrm{VTS,i}$ are the thermal photon numbers of the VTS output at the amplifier's signal frequency $\omega_\mathrm{s}$ and idler frequency $\omega_\mathrm{i}$, respectively. $G_\mathrm{sys}$ and $n_\mathrm{sys}$ are the gain and the add noise of the output line referred to the VTS, separately. $n_\mathrm{q} = 1/2$ represents quantum fluctuations. 
Note that here we assume the gain of the parametric amplifier is very large. Otherwise, the gain coefficients for $n_\mathrm{VTS,s}$ and $n_\mathrm{VTS,i}$ will be different. 

To find out $G_\mathrm{sys}$ and $n_\mathrm{sys}$, we first detune the microwave resonance from frequencies of interest, so the device
does not contribute to the calibration of $G_\mathrm{sys}$. After that, we sweep the temperature of VTS to obtain $G_\mathrm{sys}$ and $n_\mathrm{sys}$ as a function of frequency, and then fit the data to the Eq.~\ref{TWPA_calibration}. 
Given the calibrated gain results,  we can also determine the attenuation of the input line to the VTS output plane by subtracting the output line gain from the total transmission of the setup characterized with a vector network analyzer.

Due to the small attenuation between VTS and the device, the characterized device noise presented in the work should always be slightly overestimated than the actual device noise.

\section*{Appendix B: Microwave thermal bath temperature extraction} 
\setcounter{equation}{0}
\renewcommand{\theequation}{B\arabic{equation}}
\setcounter{figure}{0} 
\renewcommand{\thefigure}{B\arabic{figure}} 
To understand the microwave noise generated by optical drives in electronic-optical transducers, we need to analyze the sources of noise in these devices. To accomplish this, we will measure the thermal bath temperature that is connected to the microwave resonator through intrinsic and extrinsic channels, as illustrated in Fig.~\ref{optically_induced_microwave_noise_diagram} (a). These thermal bath temperatures represent a universal value that can be applied to other integrated EO devices with similar physical characteristics, as they are not influenced by the extraction ratio or electro-optical coupling. By studying the thermal bath temperature, we can gain a quantitative understanding of the optically induced microwave noise that can be referenced by different EO transducers, despite differences in their extraction ratio or conversion efficiency.

To measure the excessive microwave output noise, we perform an on-off characterization. First, we make sure the external magnetic field is turned off, which allows the microwave resonator to stay at its original frequency. Then we apply the optical drive, to induce excessive noise that is coupled to the microwave transmission line and the superconducting resonator. Therefore, the output noise power spectrum becomes:

\begin{equation}
P = BW\cdot G_\mathrm{sys}\cdot \hbar \omega_\mathrm{s} ( n_\mathrm{q}+n_\mathrm{sys}+\Delta n_\mathrm{sys})+BW\cdot G_\mathrm{dev}\cdot \hbar \omega_\mathrm{s}s_\mathrm{out}.
\label{seo_out_meas}
\end{equation}
Here $G_\mathrm{dev} = L\cdot G_\mathrm{sys}$, and $L$ is the attenuation between the VTS and the EO transducer. Please note that Eq.~\ref{seo_out_meas} and the following analysis are only valid when the coorperativity is much smaller than 1. Otherwise, we would need to take the loss through electro-optical effect into account. According to our previous experiment with similar configuration \cite{fu2021cavity}, the attenuation $L=91\%\pm4\%$. Thus $G_\mathrm{VTS}$ can be used as a good estimation of $G_\mathrm{dev}$ to the accuracy of 10\%. It is worth noting that because of this systematic error, the characterized device noise presented in the work should be slightly overestimated than the actual device noise.
By approximation $L=1$, we can simplify the equation to:

\begin{equation}
P = BW\cdot G_\mathrm{sys}\cdot \hbar \omega_\mathrm{s} ( n_\mathrm{q}+n_\mathrm{sys}+\Delta n_\mathrm{sys} + s_\mathrm{out}).
\label{power_noise_spectrum}
\end{equation}
Term  $\Delta n_\mathrm{sys}$ represents the potential increase in the output chain added noise after the EO transducer (see Fig.~\ref{optically_induced_microwave_noise_diagram} (b)), and $s_\mathrm{out}$ represents the power spectral density at the output of EO transducer as  

\begin{align}
s_\mathrm{out} &= a^\dagger_\mathrm{out} a_\mathrm{out}
=\mathcal{R}(\omega)n_\mathrm{e}+\mathcal{T}(\omega)n_\mathrm{i},
\label{sout}
\end{align}
where $\mathcal{T}(\omega)=\kappa_\mathrm{i}\kappa_\mathrm{e}/\big({(\kappa/2)^2+(\omega-\omega_\mathrm{0})^2}\big)$, $\mathcal{R}(\omega)=1-\mathcal{T}(\omega)$. $\omega_\mathrm{0}$, $\kappa_\mathrm{e}$ and $\kappa_\mathrm{e}$ are the angular resonant frequency, external coupling rate and intrinsic coupling rate of the superconducting resonator, respectively. $n_\mathrm{i}$ and $n_\mathrm{e}$ each denotes the power spectral density of the fields that coupled to the resonator through intrinsic and external coupling channel, separately.

\begin{figure}[h]
\includegraphics{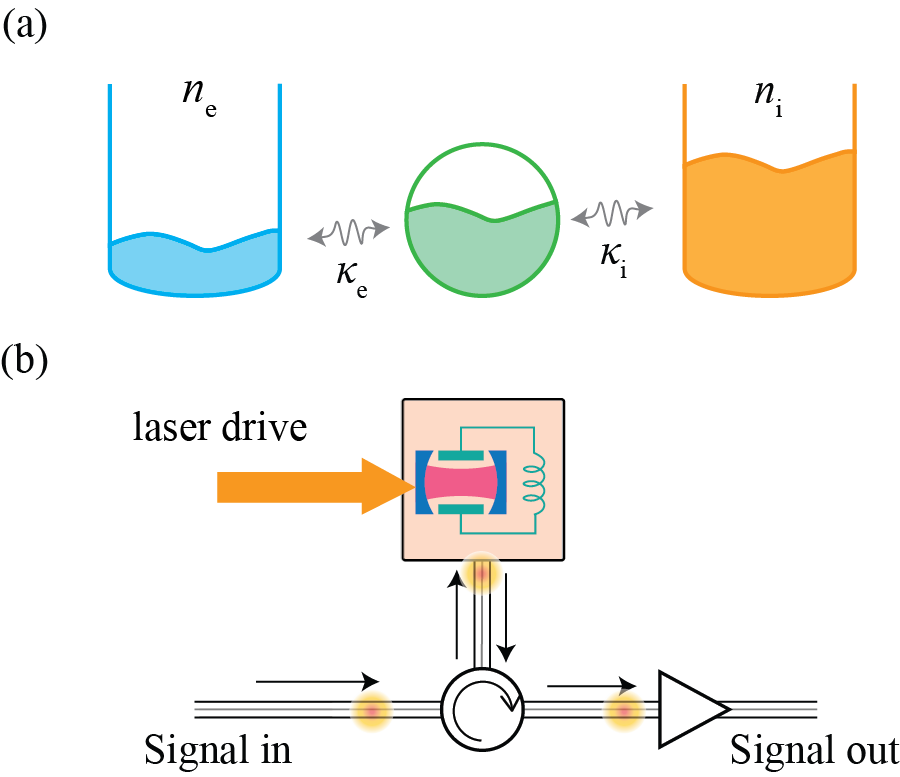}
\caption{\label{optically_induced_microwave_noise_diagram} (a) diagram of the superconducting mode being thermalized through the intrinsic and external coupling channel. (b) The possible locations where the light-induced excessive noise couples to system. }
\end{figure}

Combining Eq.~\ref{power_noise_spectrum} and Eq.~\ref{sout}, we arrive at

\begin{equation}
P = BW\cdot G_\mathrm{sys}\cdot \hbar \omega_\mathrm{s} [ n_\mathrm{q}+n_\mathrm{sys}+\Delta n_\mathrm{sys} + n_\mathrm{e}+\mathcal{T}(\omega) (n_\mathrm{i} - n_\mathrm{e})].
\end{equation}
Since in this part of the experiment, we maintain VTS at the base temperature, ususally 60~mK, its contribution to the input field of the superconducting resonator is negligible. Therefore, when the optical drive is off, we can assume $n_\mathrm{e} = 0$. Additionally, when the optical drive is off, it can be assumed that the device is thermalized to the dilution refrigerator base temperature, which is usually set to 50~mK for better cooling power. In this case, we can also assume $n_\mathrm{i} = 0$. Therefore, when the optical drive is off, the output noise spectrum can be expressed as

\begin{equation}
P_\mathrm{off} = BW\cdot G_\mathrm{sys}\cdot \hbar \omega_\mathrm{s} ( n_\mathrm{q}+n_\mathrm{sys}).
\end{equation}\par

In order to extract the light-induced microwave noise, we measure $P_\mathrm{on}$ spectrum when the optical drive is on followed by a measurement of $P_\mathrm{off}$. It is important to maintain the same gain of the output chain during the on-off measurement. Then we can extract the excess noise spectrum through

\begin{align}
s_\mathrm{dev} &= (P_\mathrm{on}-P_\mathrm{off})/(BW\cdot G_\mathrm{sys}\cdot \hbar \omega_s) \\
&=\Delta n_\mathrm{sys} + n_\mathrm{e}+\mathcal{T}(\omega) (n_\mathrm{i} - n_\mathrm{e}),
\end{align}
with the knowledge of intrinsic and external coupling rates and therefore the $\mathcal{T}(\omega)$. Note that here, we do not have the ability to separate the optically-induced noise coupled to the microwave transmission lines before and after the system, which is $n_\mathrm{e}$ and $\Delta n_\mathrm{sys}$ respectively. Therefore, we cannot determine the absolute value of $n_\mathrm{e}$ and $n_\mathrm{i}$. Instead, we are going to focus on the two parameters that can be extracted experimentally:

\begin{align}
n_\mathrm{bg} &= \Delta n_\mathrm{sys} + n_\mathrm{e},\\
n_\mathrm{i}^\mathrm{diff} &= n_\mathrm{i} - n_\mathrm{e}.
\end{align}
These experimentally variables will set the limits for the thermal bath occupancy as \begin{align}
0<&n_\mathrm{e} < n_\mathrm{bg},\\
n_\mathrm{i}^\mathrm{diff}<&n_\mathrm{i} < n_\mathrm{i}^\mathrm{diff}+n_\mathrm{bg}.
\end{align}\par

\begin{figure}[h!]
\includegraphics[width=0.45\textwidth]{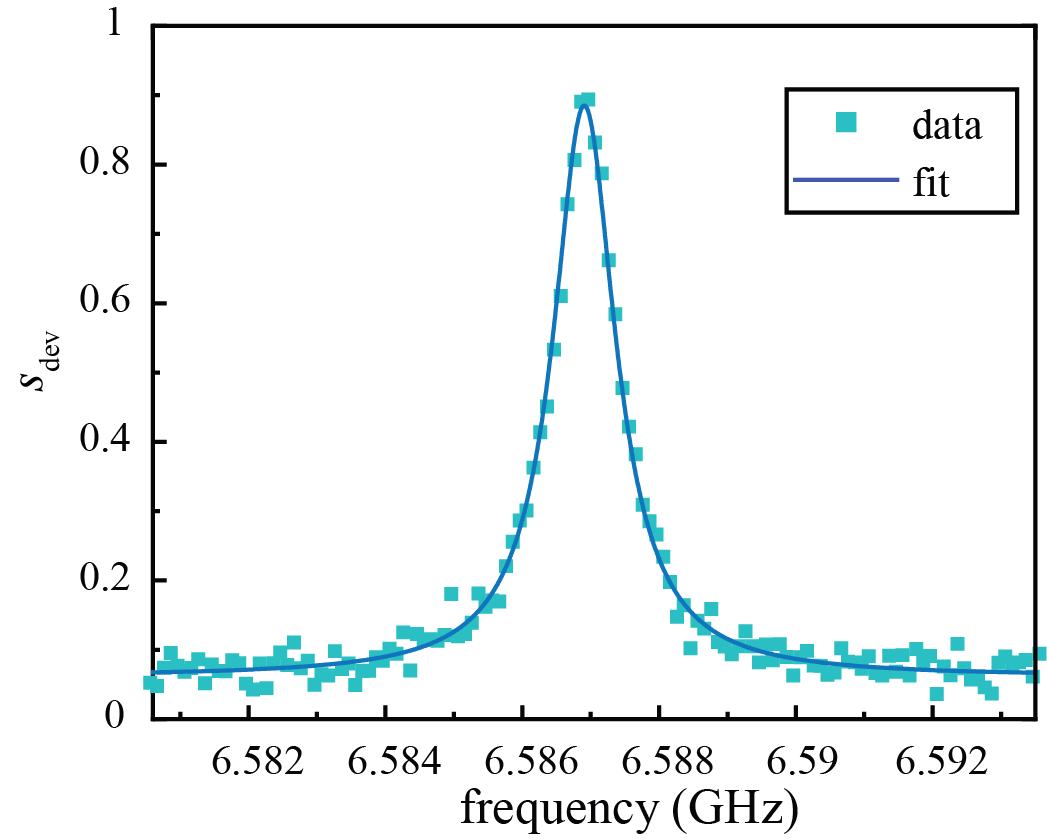}
\caption{\label{fit_eo_noise} Experimental data of $s_\mathrm{dev}$ and its fit.  }
\end{figure}

In Fig.~\ref{fit_eo_noise}, we show an example of $s_\mathrm{dev}$ curve and its fit. The model we use to fit the data is
\begin{equation}
s_\mathrm{dev} = 
n_\mathrm{bg} + \mathcal{T}(\omega) n_\mathrm{i}^\mathrm{diff}.
\end{equation}
In this model, the only free parameters are $n_\mathrm{bg}$ and $n_\mathrm{i}^\mathrm{diff}$, while the resonance frequency and coupling rates which $\mathcal{T}(\omega)$ is dependent on are extracted from coherent reflection characterization of the microwave resonator. Since the cooperativity in these devices that we study are much smaller than 1, we can ignore the broadening of the linewidth of the microwave resonator when the optical drive is on.


%

\end{document}